# Quantification of Uncertainty and Its Propagation in Seismic Velocity Structure and Earthquake Source Inversion

Ryoichiro Agata

Japan Agency for Marine-Earth Science and Technology (JAMSTEC)

3173-25 Showa-machi, Kanazawa-ku, Yokohama, Kanagawa, 236-0001, Japan

--------------------------------------

Email: agatar@jamstec.go.jp

**Abstract** *

In earthquake source inversions aimed at understanding diverse fault activities on earthquake faults using seismic observation data, uncertainties in velocity structure models are typically not considered. As a result, biases and underestimations of uncertainty can occur in source inversion. This article provides an overview of the author's efforts to address this issue by quantitatively evaluating the uncertainty in velocity structure models and appropriately accounting for its propagation into source inversion. First, the Bayesian multi-model source inversion method that can incorporate such uncertainties as probability distributions in the form of ensembles is explained. Next, a Bayesian traveltime tomography technique utilizing physics-informed neural networks (PINN) to quantify uncertainties in velocity structure models is introduced. Furthermore, the author's recent efforts to integrate these methods and apply them to hypocenter determination in the Nankai Trough region are briefly discussed. The article also outlines future prospects of source inversions considering uncertainties in velocity structure models and the anticipated role of the emerging scientific machine learning (SciML) methods such as PINN.

*Keywords*: Earthquake source inversion, uncertainty propagation, Bayesian estimation, Bayesian traveltime tomography, physics-informed neural network (PINN), scientific machine learning (SciML)

* The Abstract was specifically added for the English preprint and is not included in the original manuscript written in Japanese.

## 1. Introduction

The underground fault that hosts a huge earthquake and its surrounding areas experience a variety of fault activities, including the main shock, afterslip, small to medium-scale earthquakes, slow earthquakes, and inter-earthquake locking. Quantitatively estimating these activities using observational data such as seismic waves, crustal deformation, and tsunami data (hereafter collectively referred to as earthquake source inversion) is essential for understanding the process of strain accumulation and release in the faults, which is necessary for enhancing earthquake and tsunami hazard assessments.

When performing earthquake source inversion, it is necessary to have appropriate model assumptions about the underground medium that propagates seismic waves from the source to the observation points on the Earth's surface, such as a seismic velocity structure model. Seismic velocity structure models are estimated using data from natural earthquakes or active seismic sources where observation locations are limited to the Earth's surface, and these estimation problems are strongly nonlinear inverse problems. Due to these characteristics, even with well-constrained estimations using active seismic sources in seismic exploration, significant uncertainties can arise in the estimation results at specific depths [Korenaga et al. (2000)]. This uncertainty indicates that it is not possible to determine a single velocity structure model for a given area, and there is room for proposing many models. Nevertheless, typical source inversions are performed by selecting a single velocity structure model (for example, an average one). In practice, researchers or institutions often use further simplified models. Different research groups may propose significantly different source inversion results for the same earthquake due to differences in the assumed models. In such cases, the reliability of scientific knowledge becomes unclear [Mai et al. (2016)]. One of the key issues is that by pre-selecting a single model, the propagation of uncertainty from the velocity structure model to the source inversion is ignored, leading to underestimation of bias and uncertainty in the source inversion results [Yagi & Fukahata (2011); Duputel et al. (2014); Gesret et al. (2015); Agata et al. (2021)]. To achieve accurate and reliable source inversions, it is necessary not only to quantitatively evaluate the uncertainty of the source inversion itself [Tarantola et al. (1982); Hirata & Matsu'ura (1987); Lomax et al. (2000); Fukuda & Johnson (2008); Minson et al. (2014)] but also to properly consider the quantitative evaluation of velocity structure estimation uncertainty and its propagation into source inversion.

This article reviews the challenges related to source inversion along with relevant research and the author's previous efforts. Section 2 focuses on methods for considering model assumption uncertainties in source inversion, primarily introducing the Bayesian multi-model source inversion method proposed by the authors and providing examples of numerical experiments and applications to real data. Section 3 centers on the quantification of velocity structure model uncertainty, mainly presenting the Bayesian traveltime tomography method based on physics-informed neural networks (PINNs) [Raissi et al. (2019)] proposed by the authors, along with results from numerical experiments. Section 4 briefly introduces the authors' recent effort that integrates these approaches, implementing source inversion that considers the propagation of velocity structure model uncertainties quantified by Bayesian traveltime tomography. Section 5 provides summary and discusses future prospects. Note that Sections 2 and 3 are largely independent of each other except for the introductions, so readers can proceed to sections of interest without hindering their overall understanding of the content.

## 2. Earthquake Source Inversion Considering Uncertainty in Subsurface Structure Models

Methods that consider the impact of uncertainties in model assumptions, particularly seismic velocity structure models, on earthquake source inversion have made significant progress over the past 15 years. These studies can be broadly categorized into two main directions. One approach is to view the uncertainty in model assumptions as errors in the elastic Green's functions, which physically link source parameters and observations. In this approach, progress has been made by assuming a Gaussian distribution for the Green's function errors and setting the off-diagonal terms of the data error covariance matrix properly. The other, more direct method, involves simultaneously estimating parameters related to model assumptions, such as subsurface structure parameters, along with source parameters. These two approaches have developed in different contexts, but their theoretical connections have been explicitly demonstrated by the authors [Agata et al. (2021)]. Based on this insight, the authors proposed a Bayesian multi-model seismic source inversion method that combines the strengths of both approaches [Agata et al. (2021, 2022)]. In this section, I first review in Section 2.1 the methods involving data error covariance matrices derived from Green's function errors and those involving simultaneous estimation of source and subsurface structure parameters. I then demonstrate their close theoretical relationship within the framework of Bayesian estimation. Based on this discussion, in Section 2.2, I introduce the Bayesian multi-model source inversion method proposed by the authors, along with an example of its application to real data in Section 2.3.

**2.1 Related work**

An early study that considered off-diagonal terms of the error covariance matrix derived from Green's function errors is Yagi & Fukahata (2011), which was built upon earlier research by Yagi & Fukahata (2008) that focused on discretization errors in fault slip. This method addresses Green's function errors in seismic source process inversion using teleseismic body waves but is applicable to general linear inverse problems in earthquake source inversion. Below, I present a notation that emphasizes generality, though it may differ from the original article. In typical linear source inversion, the observation equation for the inverse analysis can be expressed as:

$$\mathbf{d} = \mathbf{H}\mathbf{m} + \mathbf{e}, \tag{1}$$

where $\mathbf{d}$ is the data vector, $\mathbf{H}$ is the matrix of theoretically computed Green's functions, and $\mathbf{m}$ is the vector of source parameters, which may include discretized seismic source processes or fault slip. The term $\mathbf{e} = \mathbf{d} - \mathbf{H}\mathbf{m}$ represents the data error component, which is assumed to follow an independent Gaussian distribution (i.e., the covariance matrix has only diagonal terms). This distribution can be considered as the likelihood function $p(\mathbf{d}|\mathbf{m})$, which is the probability density function (PDF) of $\mathbf{d}$ given $\mathbf{m}$. According to Bayes' theorem, the posterior probability distribution, combining the likelihood function with an appropriate prior distribution $p(\mathbf{m})$ for $\mathbf{m}$, is given by

$$p(\mathbf{m}|\mathbf{d}) = \kappa\, p(\mathbf{d}|\mathbf{m})p(\mathbf{m}). \tag{2}$$

where if the prior distribution is Gaussian, the posterior distribution will also be Gaussian. In this case, for a solution to the inverse problem, it is common to make a point estimation for the maximum a posteriori (MAP) solution of $p(\mathbf{m}|\mathbf{d})$. Uncertainty in the estimation can also be obtained from the analytical posterior covariance matrix. For more details, refer to textbooks such as Tarantola (2005).

However, in reality, $\mathbf{H}$ contains uncertainty due to that in the subsurface structure model. If we consider this as an error term in $\mathbf{H}$, denoted as $\delta\mathbf{H}$, then the equation becomes:

$$\begin{aligned} \mathbf{d} &= (\mathbf{H} + \delta\mathbf{H})\mathbf{m} + \mathbf{e} \\ &= \mathbf{Hm} + \delta\mathbf{Hm} + \mathbf{e}. \end{aligned} \tag{3}$$

If we consider $\delta\mathbf{Hm}$ as a new data error term, even if the components of $\delta\mathbf{H}$ follow independent Gaussian distributions, $\delta\mathbf{Hm}$ will follow a Gaussian distribution with non-zero covariance. This covariance matrix depends on $\mathbf{m}$. Therefore, the likelihood function can be written as:

$$p(\mathbf{d}|\mathbf{m}) = \kappa \exp\left[-\frac{1}{2}(\mathbf{d} - \mathbf{Hm})^{\mathsf{T}}\left(\mathbf{C}_{\mathrm{p}} + \mathbf{C}_{\mathrm{e}}\right)^{-1}(\mathbf{d} - \mathbf{Hm})\right], \tag{4}$$

where $\kappa$ is a normalization constant, $\mathbf{C}_{\mathrm{p}}$ is the covariance matrix of $\delta\mathbf{Hm}$, and $\mathbf{C}_{\mathrm{e}}$ is the covariance matrix of the Gaussian distribution that $\mathbf{e}$ follows. Given that $\mathbf{C}_{\mathrm{p}}$ has off-diagonal terms, this likelihood function is covariant. In contrast, in the basic linear source inversion based on Equation (1), the covariance matrix of the likelihood function consists only of $\mathbf{C}_{\mathrm{e}}$, which is often not assumed to have off-diagonal terms.

Given that $\mathbf{C}_{\mathrm{p}}$ depends on $\mathbf{m}$, the maximum a posteriori (MAP) solution can be obtained using non-linear optimization. The posterior covariance matrix can also be calculated analytically here. Yagi & Fukahata (2011) assumed that the standard deviation of $\delta\mathbf{H}$ is proportional to $\mathbf{H}$ and suggested to determine this scale factor using Akaike's Bayesian Information Criterion (ABIC) [Akaike, 1980]. Duputel et al. (2014)

proposed a method to theoretically compute the covariance matrix of Green's function errors based on the sensitivity of Green's functions to subsurface structure parameters. Assuming that $\mathbf{H}$ depends on the subsurface structure parameter vector $\boldsymbol{\phi}$, and using a first-order Taylor expansion around a reference point $\boldsymbol{\phi}_0$, that is, $\mathbf{H}(\boldsymbol{\phi}) \simeq \mathbf{H}(\boldsymbol{\phi}_0) + (\frac{\partial \mathbf{H}}{\partial \boldsymbol{\phi}}|_{\boldsymbol{\phi}=\boldsymbol{\phi}_0})^\top(\boldsymbol{\phi} - \boldsymbol{\phi}_0)$, the observation equation (3) becomes:

$$\mathbf{d} = \mathbf{H}(\boldsymbol{\phi}_0)\mathbf{m} + (\boldsymbol{\phi} - \boldsymbol{\phi}_0)^\top \left(\frac{\partial \mathbf{H}}{\partial \boldsymbol{\phi}}|_{\boldsymbol{\phi}=\boldsymbol{\phi}_0}\right)\mathbf{m} + \mathbf{e}. \tag{5}$$

The sensitivity matrix $\frac{\partial \mathbf{H}}{\partial \boldsymbol{\phi}}|_{\boldsymbol{\phi}=\boldsymbol{\phi}_0}$ can be derived analytically or numerically. The term $\boldsymbol{\phi} - \boldsymbol{\phi}_0$ represents the deviation from the mean subsurface structure parameter vector $\boldsymbol{\phi}_0$ and is considered a random variable to express subsurface structure uncertainty. We now introduce $\mathbf{K}_{\boldsymbol{\phi}} = \left((\frac{\partial \mathbf{H}}{\partial \boldsymbol{\phi}}|_{\boldsymbol{\phi}=\boldsymbol{\phi}_0})\mathbf{m}\right)^\top$. When we assume that this probability distribution follows a Gaussian distribution and the covariance matrix of $\boldsymbol{\phi} - \boldsymbol{\phi}_0$ is given by $\mathbf{C}_\phi$, the covariance matrix of the new error term $(\boldsymbol{\phi} - \boldsymbol{\phi}_0)^\top \left(\frac{\partial \mathbf{H}}{\partial \boldsymbol{\phi}}|_{\boldsymbol{\phi}=\boldsymbol{\phi}_0}\right)\mathbf{m} = \mathbf{K}_{\boldsymbol{\phi}}(\boldsymbol{\phi} - \boldsymbol{\phi}_0)$ is given by $\mathbf{K}_{\boldsymbol{\phi}}\mathbf{C}_\phi\mathbf{K}_{\boldsymbol{\phi}}{}^\top$. Regardless of whether the parameters are independent (i.e., whether $\mathbf{C}_\phi$ has off-diagonal elements), the sensitivity matrix is generally considered to be a dense matrix. Therefore, $\mathbf{K}_{\boldsymbol{\phi}}(\boldsymbol{\phi} - \boldsymbol{\phi}_0)$ follows a Gaussian distribution with a non-zero covariance that depends on $\mathbf{m}$. This is similar to the discussion for the method of Yagi & Fukahata (2011), requiring non-linear optimization to find the MAP solution. Duputel et al. (2014) suggested to use methods such as Markov Chain Monte Carlo (MCMC) for sampling the posterior probability distribution. Influenced by these studies, several other methods have been proposed to incorporate covariance components of data errors due to uncertainties in the model and its prediction (i.e., model prediction error) into source inversion [e.g., Hallo & Gallovič, 2016; Agata et al., 2020]. These approaches are useful because they can efficiently consider subsurface structure uncertainty without significantly increasing the number of unknowns and losing the advantages of linear inversion formulations. However, there is room for more thorough verification regarding the impact of assumptions such as the Gaussian distribution of model prediction errors and the linearization of the relationship between subsurface structure parameters and Green's functions on the analysis accuracy. Additionally, these schemes can be interpreted as methods to reduce biases and underestimations of uncertainties in source parameter estimation by implicitly using information about subsurface structure parameters contained in the data. However, they do not provide feedback on the subsurface structure parameters (although this is probably possible with some additional efforts, it has not been discussed in any of these articles).

Another approach is to estimate the parameters related to the subsurface structure model simultaneously with the source parameters (in practice, many studies estimate the fault plane parameters. However, the formulation can be considered mathematically identical to that for estimation of medium material property, so I treat them together as problems involving subsurface structure parameter here). This type of problem can be considered an estimation using the joint posterior probability distribution of both fault slip parameters and subsurface structure parameters, by adding the subsurface structure parameters to Equation (4), as

$$p(\mathbf{m}, \boldsymbol{\phi}|\mathbf{d}) = \kappa\, p(\mathbf{d}|\mathbf{m}, \boldsymbol{\phi})p(\mathbf{m}, \boldsymbol{\phi}). \tag{6}$$

Accurately determining this joint posterior probability distribution is challenging, but sampling methods such as MCMC are effective. However, because it involves two types of unknown parameter sets, the problem becomes high-dimensional, and the trade-offs between parameters can be significant, leading to a complex shape of the posterior probability distribution. As a result, achieving MCMC sampling is not easy. In practice, various joint estimation methods have been proposed to avoid such a direct approach: A method for identifying subsurface structure parameters (dip angle of a planar fault) using Akaike Bayesian Information Criterion [Fukahata & Wright (2008)], a formulation that divides the problem into linear inversion of source parameters and nonlinear inversion of subsurface structure parameters (planar fault parameters) and performs Bayesian estimation on each part [Fukuda & Johnson (2010)], simultaneously estimation of fault slip and subsurface structure parameters (such as viscosity or Poisson's ratio) using gradient-based nonlinear optimization to maximize the posterior probability [e.g., Agata et al. (2018); Puel et al. (2024)] and estimation of fault plane shape through source inversion represented by high-degree-of-freedom potential density tensors [Shimizu et al. (2021)]. Additionally, in recent years, examples have emerged where efficient parameterization and high-performance parallel computing are utilized to perform joint estimation using sampling methods such as MCMC or sequential Monte Carlo [e.g., Dutta et al. (2021); Nakao et al. (2024)].

These two distinct research directions, which have developed separately, curiously show little mutual citation. However, they are closely related mathematically [Agata et al. (2021)]. For example, let us consider eliminating the subsurface structure parameters from equation (6). By marginalizing $p(\mathbf{m}, \boldsymbol{\phi}|\mathbf{d})$ with respect to $\boldsymbol{\phi}$, we obtain

$$\begin{aligned} p(\mathbf{m}|\mathbf{d}) &= \int p(\mathbf{m}, \boldsymbol{\phi}|\mathbf{d}) d\boldsymbol{\phi} \\ &= \kappa \int p(\mathbf{d}|\mathbf{m}, \boldsymbol{\phi}) p(\mathbf{m}|\boldsymbol{\phi}) p(\boldsymbol{\phi}) d\boldsymbol{\phi}, \end{aligned} \tag{7}$$

Where the relation $p(\mathbf{m}, \boldsymbol{\phi}) = p(\mathbf{m}|\boldsymbol{\phi})p(\boldsymbol{\phi})$ is used. If such marginalization can be performed in advance, it becomes possible to estimate only the fault slip parameters while still accounting for the variability of the subsurface structure expressed by $p(\boldsymbol{\phi})$. The integral required for marginalization is generally not analytically computable, but can be computed if, for example, the integrand is a Gaussian function. The posterior PDF formulated by Duputel et al. (2014) can actually be shown to be equivalent to the marginalized posterior PDF of the subsurface structure parameters under the following conditions: The prior distribution of the subsurface structure parameters $p(\boldsymbol{\phi})$ is a Gaussian distribution, the likelihood function is also a Gaussian distribution, and the relationship between the Green's function and the subsurface structure parameters can be linearly approximated. We rearrange the right-hand side in Equation (7) as $\kappa\, p(\mathbf{m}) \int p(\mathbf{d}|\mathbf{m}, \boldsymbol{\phi}) p(\boldsymbol{\phi}) d\boldsymbol{\phi}$ with $p(\mathbf{m}|\boldsymbol{\phi}) = p(\mathbf{m})$ for simplicity. Given $p(\mathbf{d}|\mathbf{m}, \boldsymbol{\phi}) = \kappa \exp\left[-\frac{1}{2}(\mathbf{d} - \mathbf{H}(\boldsymbol{\phi})\mathbf{m})^\top {\mathbf{C}_\mathrm{e}}^{-1}(\mathbf{d} - \mathbf{H}(\boldsymbol{\phi})\mathbf{m})\right]$, $p(\boldsymbol{\phi}) = \kappa \exp\left[-\frac{1}{2}(\boldsymbol{\phi} - \boldsymbol{\phi}_0)^\top {\mathbf{C}_\phi}^{-1}(\boldsymbol{\phi} - \boldsymbol{\phi}_0)\right]$ and the first-order Taylor expansion of $\mathbf{H}(\boldsymbol{\phi})$ around a reference point $\boldsymbol{\phi}_0$, this integral becomes

$$\int p(\mathbf{d}|\mathbf{m},\boldsymbol{\phi})p(\boldsymbol{\phi})d\boldsymbol{\phi} = \kappa\int \exp\left[-\frac{1}{2}\left(\mathbf{u}^\top\mathbf{C}_\mathrm{e}{}^{-1}\mathbf{u}\right) - \frac{1}{2}(\boldsymbol{\phi}-\boldsymbol{\phi}_0)^\top\mathbf{C}_\phi{}^{-1}(\boldsymbol{\phi}-\boldsymbol{\phi}_0)\right]d\boldsymbol{\phi}, \tag{8}$$

where $\mathbf{u} = \mathbf{d} - \mathbf{H}(\boldsymbol{\phi}_0)\mathbf{m} - \mathbf{K}_{\boldsymbol{\phi}}(\boldsymbol{\phi}-\boldsymbol{\phi}_0)$ By organizing this and applying the formula for integrating Gaussian functions, the same data error covariance matrix $\mathbf{K}_{\boldsymbol{\phi}}\mathbf{C}_\phi\mathbf{K}_{\boldsymbol{\phi}}{}^\top$ as derived by Duputel et al. (2014) can be obtained. Yagi & Fukahata (2011) can be interpreted as determining the error scale in a data-driven manner without tracing back to the origin of Green's function errors in their similar formulation.

**2.2 Bayesian multi-model source inversion**

Agata et al. (2021) noted that if an ensemble model representing the uncertainty in subsurface structure is available, we can interpret the ensemble is the samples from $p(\boldsymbol{\phi})$. Then, the marginalization of Equation (7) can be approximately computed using Monte Carlo integration ( $\int f(x)p(x)dx \approx \frac{1}{N}\sum_{n=1}^{N} f\left(x^{(n)}\right)$ , where $x^{(n)} \sim p(x)$ ) without introducing assumptions of Gaussian distributions or linearity, as

$$\begin{aligned} p(\mathbf{m}|\mathbf{d}) &= \kappa\int p(\mathbf{d}|\mathbf{m},\boldsymbol{\phi})p(\mathbf{m}|\boldsymbol{\phi})p(\boldsymbol{\phi})d\boldsymbol{\phi} \\ &\simeq \kappa\frac{1}{N}\sum_{n=1}^{N} p\left(\mathbf{d}\middle|\mathbf{m},\boldsymbol{\phi}^{(n)}\right)p\left(\mathbf{m}\middle|\boldsymbol{\phi}^{(n)}\right). \end{aligned} \tag{9}$$

In many problems, $p\left(\mathbf{m}\middle|\boldsymbol{\phi}^{(n)}\right) = p(\mathbf{m})$. This means that, if an ensemble of subsurface structure models is available, it is possible to sample from the marginalized posterior distribution of source parameters based on this equation. The idea of using Monte Carlo integration for marginalizing subsurface structure parameters has already been demonstrated in the context of induced microseismicity by Poliannikov et al. (2014) and Gesret et al. (2015). Agata et al. (2021) further showed that an approximate representation of the posterior distribution of subsurface structure parameters $p(\boldsymbol{\phi}|\mathbf{d})$ can be obtained as a post process of the fault slip parameters derived from this method. Simply put, instead of marginalizing with respect to $\boldsymbol{\phi}$ as shown in equation (7), we consider marginalization with respect to $\mathbf{m}$. After appropriately transforming the integrant, we evaluate the posterior probability using Monte Carlo integration with samples of the fault slip parameters. Given that this approach of Agata et al. (2021) shares a similar theoretical background with Bayesian multi-model estimation techniques used in meteorology, such as data assimilation using multiple weather models (Tebaldi & Knutti, 2007), also known as Bayesian model averaging (Raftery et al., 1997), Agata et al. (2022) redefined this method as Bayesian multi-model fault slip estimation.

The key feature of Bayesian multi-model source inversion is that it can account for the non-Gaussian prior probability (i.e., uncertainty) of subsurface structure parameters and their nonlinear relationship with the Green's function without increasing the number of unknowns in the estimation. Despite this, it is still possible to estimate subsurface structure parameters, similar to a joint estimation framework. Note that while the number of unknowns does not increase, the computational cost for Monte Carlo integration does rise. Additionally, if the prior distribution of subsurface structure parameters is not sufficiently close to the posterior distribution, a large number of samples will be required.

### 2.3 Application to Fault Slip Estimation for the Long-Term Slow Slip in the Bungo Channel

In the plate boundary beneath the Bungo Channel in southwestern Japan, long-term slow slip events (L-SSEs) with durations of about 6 months to 1 year and recurrence intervals of 6 to 8 years, and magnitudes of Mw 6 to 7, repeatedly occur almost in the same location. During L-SSE occurrences, it is known that the number of deep tectonic tremors increases on the down-dip side of the main rupture zone [Hirose et al. (2010)]. Agata et al. (2022) used the Bayesian multi-model fault slip estimation method to estimate the fault slip distribution of L-SSEs that occurred around 2010 and in 2018, based on geodetic observation data, with the aim of comparing the fault slip distribution with the distribution of deep tectonic tremors. By appropriately considering the uncertainty in subsurface structure models, which is assumed to be a major source of model prediction errors included in the data error term, it is expected that the instability in the inversion analysis of fault slip distribution can be reduced. This approach aims to enable estimation without imposing strong prior constraints, such as smoothing, which are often applied to fault slip distributions. Agata et al. (2022) attempted to compare the results of estimating the slip distributions for the two L-SSEs in 2010 and 2018, excluding strong prior constraints, with the distribution of deep tectonic tremors.

To express the uncertainty in the subsurface structure around the rupture zone as an ensemble approximating the prior distribution, 2,000 two-layer structure and fault geometry models were created for the elastic structure and plate boundary geometry of the area. The plate boundary geometry was considered as a weighted average model based on the Iwasaki model by Iwasaki et al. (2015), Slab2 [Hayes et al. (2018)], and the Nakanishi model by Nakanishi et al. (2018). The weights were set to a Dirichlet distribution with $\alpha$=1, which corresponds to completely random probabilities for each weight (Fig. 1(a)). The elastic structure was defined by setting the thickness of the first layer as the continental crust thickness and the average physical properties of the continental crust and mantle as the properties of the first and second layers, respectively, based on Japan Integrated Velocity Structure Model (JIVSM) [Koketsu et al. (2009, 2012)] at 2,000 randomly selected points within the horizontal area of interest. The surface displacement Green's function for fault slip in a two-layer elastic structure was calculated using EDGRN/EDCMP [Wang et al. (2003)]. For the crustal deformation data of L-SSEs, the local tectonic component of crustal deformation extracted from GEONET data by Seshimo and Yoshioka (2022) was used. The parameters for the fault slip distribution were set as discretized parameters using bilinear interpolation functions on a grid with 16 km intervals in the area with the color contour shown in Fig. 2(a). The slip direction was assumed to be uniform across all small faults, and this slip direction was also estimated from the data. Instead of a prior distribution equivalent to smoothing commonly used in fault slip distribution estimation, a weak prior distribution given by a uniform distribution was applied to the slip amount at each small fault.

Here, I present the estimation results for the event around 2010. Compared to the slip distribution obtained with prior smoothing constraints assuming a single underground structure [e.g., Yoshioka et al. (2015)], the extent of the region with large slip in the down-dip direction in the mean model of the posterior probability distribution of the fault slip distribution obtained in this study was found to be narrower (Fig. 2(a)). The Coulomb failure stress change (ΔCFS) on the profile AB was calculated and compared with an independently estimated result based on smoothing constraints and a single underground structure model.It is evident that our estimation results better correspond to the distribution of deep tectonic tremors during this event [Maeda & Obara (2009); Obara (2010); Kano et al. (2018)] (Fig. 2(b)). By appropriately considering the uncertainty in subsurface structure models and reducing the instability caused by model prediction errors in the inverse analysis, this study provides an example of how fault slip distribution estimation without strong prior information can offer important insights into the physical relationships between multiple earthquake events. Additionally, through fault slip distribution estimation, the posterior probability distribution of subsurface structure models was also obtained. This is one of the advantages of Bayesian multi-model estimation. The elastic structure was not significantly updated from the prior distribution. On the other hand, for the fault shape model, it became evident that a model similar to the average of both Iwasaki and Nakanishi models and Slab2 had a higher probability, indicating feedback from crustal deformation data (Fig. 1(b)). This result is consistent with previous studies suggesting that fault slip estimation using crustal deformation data is more sensitive to fault shape [e.g., Lindsey and Fialko (2013)]. Agata et al. (2022) also explored the potential of Bayesian multi-model estimation by estimating the fault slip distributions for the 2010 and 2018 L-SSEs while sequentially updating the probability distribution of subsurface structure models. For more details, please refer to the article.

## 3. Quantification of Velocity Structure Estimation Uncertainty

The framework of Bayesian multi-model source inversion enables seismic source inversion considering the uncertainty in subsurface structures represented as an ensemble. However, as shown in the example in section 2.3, the method for constructing the ensemble of subsurface structures is not straightforward. A suboptimal approach is to assume a larger uncertainty for a roughly guessed mean values. A similar approach has also been taken in applications of Duputel et al. (2014)'s method to real data [e.g., Gombert et al. (2018); Ragon et al. (2019)]. On the other hand, Gesret et al. (2015) used a velocity structure ensemble model estimated by Bayesian traveltime tomography using active seismic exploration data in a method similar to Bayesian multi-model estimation for hypocenter determination, although it was a simpler tomographic setting with fewer parameters. In this way, a more natural approach to implementing Bayesian multi-model source inversion is to perform velocity structure estimation itself, such as traveltime tomography, using Bayesian estimation and use the samples obtained there as inputs to the multi-model estimation. Bayesian traveltime tomography generally becomes an estimation problem with a high-dimensional parameter space for velocity structures and has been considered a challenging issue. However, with the recent development of numerical calculation techniques, it is now becoming sufficiently feasible. Here, I explain an overview of one of the current mainstream methods in Bayesian traveltime tomography and discuss the issues when using the velocity structure samples obtained from this method as input to Bayesian multi-model source inversion. I then introduce a new Bayesian traveltime tomography method based on physics-informed neural networks (PINN) proposed by the authors to overcome these issues.

### 3.1 Related work

In first-arrival refraction traveltime tomography using active seismic data, the most established method for quantifying uncertainty is Monte Carlo uncertainty analysis. This involves performing multiple tomographic inversions under random data errors and initial models to examine the mean and variability of the solutions [Zhang & Toksoz (1998); Korenaga et al. (2000)]. This method has been widely adopted because it can use the same computer program for usual deterministic tomographic analyses where the initial model is fixed.

For more rigorous Bayesian estimation methods, the trans-dimensional Bayesian tomography method by Bodin & Sambridge (2009) is well-known in the context of surface wave tomography. This method uses a sampling technique called reversible jump MCMC [Green (1995)], which can randomly sample the number of parameters. As its name suggests, it parameterizes velocity structures with different dimensions for each sample. Voronoi tessellation is widely used for discretization in this parameterization. The low-dimensional parametrization using Voronoi tessellation significantly reduces the number of parameters compared to sampling velocity structures on a dense grid, which is one of the key factors that make this method easier to sample. Although each individual sample includes unnatural discontinuities due to Voronoi cells, these do not pose problems because the focus of this method is on statistical quantities such as the mean and standard deviation of the ensemble. Additionally, the introduction of Voronoi cells eliminates the need for explicit smoothing constraints. This characteristic has led to the method's wider application in local earthquake wave tomography using natural earthquakes [Piana Agostinetti et al. (2015)] and first-arrival traveltime tomography [Ryberg & Haberland (2018); Ryberg et al. (2023)]. This is a well-designed method for the purpose of quantifying uncertainty in traveltime tomography. However, from the perspective of creating an input ensemble velocity structure model for Bayesian multi-model source inversion, it is impractical to apply velocity structure models that include discontinuous and low-dimensional parametrization due to Voronoi cells to subsequent analyses. These velocity structure models have only the minimum degrees of freedom necessary to explain the data used for their estimation, and therefore they cannot be used as models to explain other data. Therefore, other methods need to be considered.

The success of trans-dimensional Bayesian tomography highlights that the method of discretizing velocity structure parameters is a key element. On the other hand, PINN [Raissi et al. (2019)], which have gained attention in recent years, have brought a new direction for parameterization in inverse problems based on partial differential equations (PDEs), such as tomographic analysis. Unlike numerical methods based on grid or mesh-based discretization of solutions, such as finite difference or finite element methods, PINNs use neural networks (NNs) to directly represent the solution as a continuous function. The solution is obtained by optimizing this function to minimize the residual of the governing PDE at appropriately set evaluation points. Originally, PINN was proposed as a framework for forward analysis, solving PDEs under initial and boundary conditions. First, consider the forward analysis of the eikonal equation, which forms the basis of traveltime tomography:

$$|\nabla T(\mathbf{x}, \mathbf{x}_\mathrm{S})|^2 = \frac{1}{v^2(\mathbf{x})}, \forall \mathbf{x} \in \Omega \tag{10}$$

$$T(\mathbf{x}_\mathrm{S}, \mathbf{x}_\mathrm{S}) = 0, \tag{11}$$

where, $\mathbf{x}$ and $\mathbf{x}_\mathrm{S}$ are the receiver and source coordinates, $T(\mathbf{x}, \mathbf{x}_\mathrm{S})$ is the travel time between the receiver and source, and $v(\mathbf{x})$ is the velocity structure. Equation (11) represents the source condition that must be satisfied (the travel time from the source to itself is 0). To automatically satisfy the source condition and to mitigate the singularity at the source, a factorization decomposition $T = T_0\tau$, $T_0 = |\mathbf{x} - \mathbf{x}_\mathrm{S}|$ is performed and an NN function to approximate $\tau$ is introduced [Smith et al. (2021), Waheed et al. (2021a)] (Fig. 3(a)). The authors proposed approximating $\frac{1}{\tau}$ using an NN function [Agata et al. (2023)]. Following this approach, I introduce the approximation function $f_{\tau^{-1}}(\mathbf{x}, \mathbf{x}_\mathrm{S}; \boldsymbol{\theta}_T)$. Here, $\boldsymbol{\theta}_T$ represents the weight parameters of the NN. Therefore, the approximate function for the travel time itself is given by $f_T(\mathbf{x}, \mathbf{x}_\mathrm{S}; \boldsymbol{\theta}_T) = T_0/f_{\tau^{-1}}(\mathbf{x}, \mathbf{x}_\mathrm{S}; \boldsymbol{\theta}_T)$. To approximate $T(\mathbf{x}, \mathbf{x}_\mathrm{S})$, we introduce a function $f_T(\mathbf{x}, \mathbf{x}_\mathrm{S}; \boldsymbol{\theta}_T)$ based on a neural network (NN) [Smith et al. (2021); Waheed et al. (2021a)] (Fig. 3(a)), where $\boldsymbol{\theta}_T$ is the vector for weight parameters. Random sample pairs are generated for $\mathbf{x}$ from the domain of interest where we solve the differential equation with respect to and for $\mathbf{x}_\mathrm{S}$ from the region where we want to evaluate the sources. The samples for the former are particularly called collocation points, denoted as $\mathbf{x}_\mathrm{c}$. By substituting $f_T(\mathbf{x}_\mathrm{c}, \mathbf{x}_\mathrm{s}; \boldsymbol{\theta}_T)$ into the eikonal equation, we calculate the equation residual for each random sample pair. The approximate solution can be obtained by minimizing a loss function consisting of the sum of squared residuals of (the square of) Equation (10), for example,

$$L = \sum_{i=1}^{N_c} \left( \left| \nabla f_T \left( \mathbf{x}_\mathrm{c}^{(i)}, \mathbf{x}_\mathrm{s}^{(i)}; \boldsymbol{\theta}_T \right) \right| - \frac{1}{v\left(\mathbf{x}_\mathrm{c}^{(i)}\right)} \right)^2, \tag{12}$$

and learning the weights of the NN. Here, $N_\mathrm{c}$ is the number of random sample pairs, and $i$ is their index. The derivative $\nabla f_T \left( \mathbf{x}_\mathrm{c}^{(i)}, \mathbf{x}_\mathrm{s}^{(i)}; \boldsymbol{\theta}_T \right)$ in the residual term can be computed using automatic differentiation. In the case of solving inverse problems, the misfit with data is added to the loss function for formulation. The parameters to be estimated can also be modeled using an NN. For example, in traveltime tomography, we introduce another NN function $f_v(\mathbf{x}; \boldsymbol{\theta}_v)$ to approximate the velocity structure $v(\mathbf{x})$ [Waheed et al. (2021b); Chen et al. (2022)] (Fig. 3(b)), where $\boldsymbol{\theta}_v$ is the vector for weight parameters. The loss function can be expressed using observed travel times $T_\mathrm{obs}$ and observation point coordinates $\mathbf{x}_\mathrm{r}$ as:

$$L = \alpha_1 \sum_{i=1}^{N_c} \left( \left| \nabla f_T \left( \mathbf{x}_\mathrm{c}^{(i)}, \mathbf{x}_\mathrm{s}^{(i)}; \boldsymbol{\theta}_T \right) \right| - \frac{1}{f_v\left(\mathbf{x}_\mathrm{c}^{(i)}; \boldsymbol{\theta}_v\right)} \right)^2 + \alpha_2 \sum_{i=1}^{N_T} \left( T_\mathrm{obs}^{(i)} - f_T \left( \mathbf{x}_\mathrm{r}^{(i)}, \mathbf{x}_\mathrm{s}^{(i)}; \boldsymbol{\theta}_T \right) \right)^2, \tag{13}$$

where $N_T$ is the number of observation points, and $\alpha_1$ and $\alpha_2$ are appropriately chosen weights. Note that $f_v(\mathbf{x}; \boldsymbol{\theta}_v)$ is introduced instead of $v(\mathbf{x})$ in the equation residual term. By simultaneously learning the weights of the two NNs to minimize this loss function, traveltime tomography can be conducted theoretically. PINN-based forward and inversion modelings are fully mesh-free and can be applied to repeated evaluations at a reduced number of random points (i.e., mini-batch training). If a PINN-based approach can be successfully applied to Bayesian tomography, it has the potential to free us from the challenges of high-dimensional Bayesian estimation due to the detailed grid required for unknown parameters, as well as the unnatural discontinuous structures introduced by low-dimensional parameterizations in the trans-dimensional Bayesian tomography.

On the other hand, extending the forward and inverse analyses using PINN to Bayesian estimation (Bayesian PINN or B-PINN [Yang et al. (2021)]) is not trivial, and there are few examples of successful implementation. The approach that primarily performs Bayesian estimation on NN weight parameters and calculates probability distribution the prediction of NN follows (predictive PDF) is known as a Bayesian neural network (BNN). In BNN, the method that most accurately estimate the true posterior and predictive distribution is considered to be Hamiltonian Monte Carlo (HMC) [Duane et al. (1987)], a type of MCMC that uses the gradient of the target probability density with respect to the unknown parameters to improve sampling efficiency. The method of applying HMC to PINN [Yang et al. (2021)] is a widely recognized B-PINN approach, but its application to real-world problems is limited to small-scale problems [Linka et al. (2022)]. The primary factors that make it difficult to scale up HMC-based B-PINN is the high multimodality of the posterior distribution in the NN parameter space. This multimodality is a consequence of the nonlinear mapping from the input to output enabled by the over-parameterized representation in NNs realizing their high function approximation capability, which allows for similar outputs to be generated by different combinations of weight parameters. Other technical factors include the difficulty in parallelizing the method due to the need for many sequential evaluations and the fact that It is necessary to process all the data for a single sample update (i.e., does not support mini-batch training) and is therefore unsuitable for processing large-scale data. Additionally, in Bayesian estimation in weight space, it is necessary to introduce a prior distribution $p(\boldsymbol{\theta})$ over the weight parameters. Setting this prior distribution in a physically meaningful way is also not straightforward. In summary, while PINN-based tomography methods may have plausible features in terms of parameterization of velocity structures, the method for extending to Bayesian estimation was not straightforward.

### 3.2 PINN-based Bayesian Traveltime Tomography

Considering the aforementioned issues with HMC in B-PINN, the authors proposed a method for PINN-based Bayesian traveltime tomography using function-space particle-based variational inference [Wang et al. (2019)] as an alternative Bayesian estimation technique [Agata et al. (2023)]. I explain how the proposed method addresses the shortcomings of the HMC-based B-PINN approach (also see the summary in Table 1). Particle-based variational inference (ParVI), represented by Stein variational gradient descent (SVGD) [Liu & Wang (2016)], is a method for approximating distributions using samples (particles), similar to sampling methods like MCMC. However, unlike MCMC, which generates samples sequentially through a Markov chain, particle-based variational inference optimizes and approaches the target distribution by interacting a set of particles that are generated in advance. The algorithm itself is simple, and for SVGD used here, the update rule for the *i*-th particle's parameter $\boldsymbol{\theta}_i$ among the particles to approximate $p(\boldsymbol{\theta})$ is given by the following equation.

$$\boldsymbol{\theta}_i = \boldsymbol{\theta}_i + \eta \boldsymbol{\phi}(\boldsymbol{\theta}_i),$$

$$\boldsymbol{\phi}(\boldsymbol{\theta}_i) = \frac{1}{n}\sum_{j=1}^{n} \left\{ k(\boldsymbol{\theta}_j, \boldsymbol{\theta}_i) \nabla_{\boldsymbol{\theta}_j} \log p(\boldsymbol{\theta}_j) + \nabla_{\boldsymbol{\theta}_j} k(\boldsymbol{\theta}_j, \boldsymbol{\theta}_i) \right\}, \quad (14)$$

where $n$ is the number of particles and $\eta$ is the step size in the current step. The step size can be determined using various adaptive optimizers and $k(\boldsymbol{\theta},\cdot)$ represents a positive definite kernel, for which we adopt a radial basis function (RBF) kernel with bandwidth determined by the median heuristic as in previous studies. Note that in this context, $p(\boldsymbol{\theta})$ generally refers to the target probability distribution rather than the prior distribution of $\boldsymbol{\theta}$. The first term within {} on the right-hand side represents the smoothed gradient of the target distribution's logarithm, while the second term represents the interaction between each particle. The latter is also known as the repulsive force term and has the effect of preventing particles from clustering together by repelling each other when they get too close during gradient descent. This enhances the distribution approximation efficiency of SVGD, and it is known that a good approximation can often be achieved with just a few hundred particles in many problems. Since the gradient calculation for the first term can be computed independently for each particle at each optimization step, the algorithm is highly parallelizable. Additionally, because the method involves optimizing to find a probability distribution, it is also compatible with mini-batch learning using stochastic gradient descent or similar techniques. However, Wang et al. (2019) pointed out that the performance of particle-based variational inference degrades for BNNs of a certain size or larger, likely due to the increased multimodality of the probability distribution in weight parameter space. Instead, Wang et al. (2019) proposed function-space particle-based variational inference, which performs optimization in the output space (function space) of the NN. Specifically, considering an NN function $f(\mathbf{x};\boldsymbol{\theta})$ that takes data $\mathbf{x}$ as input, and letting $\mathbf{f} = [f_i] = [f(\mathbf{x}_i;\boldsymbol{\theta})]$ be the vector of predictions for multiple data points, the update vector in Equation (14) is revised to:

$$\boldsymbol{\phi}(\boldsymbol{\theta}_i) = \frac{1}{n}\frac{\partial \mathbf{f}_i}{\partial \boldsymbol{\theta}_i}^{\top} \sum_{j=1}^{n} \left\{ k(\mathbf{f}_j, \mathbf{f}_i) \nabla_{\mathbf{f}_j} \log p(\mathbf{f}_j) + \nabla_{\mathbf{f}_j} k(\mathbf{f}_j, \mathbf{f}_i) \right\}. \tag{15}$$

This equation calculates the update quantity in the function space of $\mathbf{f}$, and then converts it to the weight parameter space using the Jacobian matrix. The key point of this method is that the repulsive force term is calculated in the function space. Another notable aspect is that the target distribution for which the gradient is calculated is $p(\mathbf{f}_j)$, i.e., the distribution of the NN's output values. This is because the criterion for "closeness" between particles, which is necessary to compute effective repulsive forces, becomes meaningful in function (physical) space rather than in the over-parameterized weight parameter space. In the weight parameter space, simple metrics like Euclidean distance lose their physical significance due to the over-parameterization. By working in function space, where the mappings and their physical interpretations are clearer, we can more effectively define and compute these repulsive forces. This opens up a way to directly introduce physically meaningful prior distributions. By applying this function space-based method, it is expected that accurate B-PINN based on particle-based variational inference can be achieved. A point to consider in this context is that two NNs are handled in PINN-based Bayesian tomography: one for the velocity structure and another for travel times. Our focus here is on the uncertainty of the velocity structure, while there is no need for uncertainty quantification in travel times. Therefore, the function-space particle-based variational inference algorithm given by Equation (15) is applied only to the velocity structure NN. Specifically, Equation (15) is rewritten with respect to the posterior probability distribution of the velocity structure $\mathbf{v} = [f(\mathbf{x}_i;\boldsymbol{\theta}^v)]_i$ output by the velocity structure NN as

$$\boldsymbol{\phi}(\boldsymbol{\theta}_{v\,i}) = \frac{1}{n}\frac{\partial \mathbf{v}_i}{\partial \boldsymbol{\theta}_{v\,i}}^{\top}\sum_{j=1}^{n}\left\{k(\mathbf{v}_j,\mathbf{v}_i)\nabla_{\mathbf{v}_j}\log p(\mathbf{v}_j|\mathbf{d}) + \nabla_{\mathbf{v}_j}k(\mathbf{v}_j,\mathbf{v}_i)\right\} \tag{16}$$

where the target distribution is rewritten as the posterior probability distribution of the velocity structure output by the $j$-th particle. The posterior distribution that serves as the target is composed as $p(\mathbf{v}|\mathbf{d}) \propto p(\mathbf{d}|\mathbf{v})p(\mathbf{v})$. The probability distribution that $\mathbf{v}$ follows is obtained directly as a posterior distribution, not as a predictive distribution based on the posterior probability distribution of the weights. It is important to emphasize again that this allows for incorporating a physically meaningful prior distribution $p(\mathbf{v})$ of the velocity structure. The issue here is the computation of the gradient of the posterior probability with respect to the velocity structure, $\nabla_{\mathbf{v}_j}\log\, p(\mathbf{v}_j|\mathbf{d})$. The likelihood function $p(\mathbf{d}|\mathbf{v})$, which forms the posterior distribution, is actually calculated using only the observed data and the output of the travel time NN function $f_T(\mathbf{x}_c,\mathbf{x}_s;\boldsymbol{\theta}_T)$, which is not an explicit function of $\mathbf{v}$. Therefore, it is not possible to directly compute $\nabla_{\mathbf{v}}\log\, p(\mathbf{d}|\mathbf{v})$, i.e., the derivative of the likelihood function with respect to $\mathbf{v}$, using automatic differentiation. To address this, Agata et al. (2023) proposed using the discrete adjoint method to compute this gradient. To avoid making the discussion overly complex, I omit the details here. Interested readers are referred to the article.

The method development described above enables Bayesian estimation in the function space of NNs for PINN-based tomography, offering high approximation accuracy of the posterior PDF, good parallelism, compatibility with mini-batch learning for processing large-scale data, and the ability to introduce physically meaningful prior distributions. Furthermore, representing the velocity structure as a continuous function based on NNs is expected to be a more suitable method for creating input ensemble velocity structure models for Bayesian multi-model earthquake source inversion.

### 3.3 Numerical Experiment Examples

Here, I introduce part of the numerical experiment content presented by Agata et al. (2023). In their study, the method was first validated through a one-dimensional Bayesian tomography numerical experiment assuming a linear-Gaussian distribution for which an analytical solution exists. Synthetic travel time data were computed based on the specified velocity structure and source-receiver configurations, and the Bayesian tomography using function-space particle-based variational inference with 256 particles was applied. For various settings of the correlation length of the radial basis function, which is the kernel function of the Gaussian process used as the prior distribution, the proposed method consistently obtained posterior probability distributions that closely matched the quasi-analytical solution. When a similar problem was solved using a conventional particle-based variational inference Bayesian tomography method without considering the function space, an approach essentially identical to the method proposed by Gou et al. (2023), the estimated uncertainty (standard deviation) of the velocity structure was significantly underestimated. Additionally, because the estimation was performed in the weight space, it was difficult to set physically meaningful prior distributions. These results confirm the advantages of the proposed method, which introduces Bayesian estimation in the function space of NNs.

Agata et al. (2023) further conducted a numerical experiment on 2D first-arrival refraction traveltime tomography, simulating active seismic exploration. The objective was to verify whether reasonable solutions could be obtained and whether the velocity structure samples obtained would have properties suitable as inputs for Bayesian multi-model earthquake source inversion. Based on the setting shown in Fig. 4(a), synthetic travel time data were calculated for a relatively simple velocity structure with 20 sources and 96 receivers uniformly spaced between horizontal distances of 5 km to 195 km at the surface. The Bayesian tomography using function-space particle-based variational inference with 512 particles was applied under a prior distribution based on a Gaussian process. The mean (Fig. 4(b)) and standard deviation (Fig. 4(c)) of the approximated posterior distribution were reasonable, showing good agreement with the true values and the specified prior distribution. As shown in Fig. 5, the velocity structure samples obtained exhibit continuous functions, unlike the unnatural discontinuous structures seen in low-dimensional grid-based velocity structure samples from previous Bayesian tomography methods. These results confirm that the proposed method is a promising foundational approach for performing velocity structure estimation based on Bayesian inference and using the obtained samples as inputs for Bayesian multi-model estimation.

## 4. Quantifying the Propagation of Velocity Structure Model Uncertainty in Earthquake Source Inversion:

Through the studies and developments presented in Sections 2 and 3, the author aims to integrate both approaches to perform earthquake source inversion that considers appropriately quantified uncertainties in seismic velocity structures. In this section, I briefly introduce the authors' recent preliminary efforts towards this integration, as described in Agata et al. (2025).

On April 1, 2016, an earthquake with a magnitude of Mw5.9 occurred off the southeastern coast of Mie Prefecture. Initially, there was debate over whether this earthquake originated at the plate boundary. Subsequently, studies using the structural information for hypocenter determination and comparisons with reflection profiles from seismic reflection surveys suggested that the earthquake likely occurred at or near the plate boundary [Wallace et al. (2016); Nakano et al. (2018)]. However, uncertainties in the subsurface structure models, which were not considered in these analyses, are thought to have affected both the hypocenter determination and the depth of the reflection profiles used for comparison. Re-examining the earthquake's hypocenter location by accounting for these uncertainties is desirable. Agata et al. (2025) performed an ensemble estimation of seismic velocity structures beneath a survey line from marine seismic exploration off the southeastern coast of Mie Prefecture using a PINN-based Bayesian traveltime tomography method [Agata et al. (2023)]. The estimated velocity structure ensemble was then used as input for a Bayesian multi-model hypocenter determination of the earthquake. For the velocity structure ensemble estimation, first-arrival travel times from refraction survey line KI03, which passes near the estimated epicenter of the Mw5.9 earthquake, were utilized to estimate a 2D P-wave velocity structure ensemble. In the Bayesian multi-model hypocenter determination, P-wave travel time data from all nine observation points at the nodes D and E of DONET [Kaneda et al. (2015); Aoi et al. (2020)] were used [Nakano et al. (2018)]. Given that the length of the modeling domain perpendicular to the trench axis was not very large, i.e., only 36 km, the estimated 2D structures were extended in this direction to create a 3D velocity structure (so-called "2.5D structure") for use in travel time calculations for hypocenter determination. PINN was also used for these travel time calculations. Specifically, by pre-training PINNs to learn travel time solutions for various potential hypocenter locations within the search range for each velocity structure, it became possible to rapidly evaluate numerous travel times required during the Bayesian multi-model estimation.

The estimated hypocenter location, when accounting for subsurface structural uncertainties, was found to be deeper in the mean model with greater uncertainty compared to cases where these uncertainties were not considered. This implies that the bias and underestimation of uncertainty introduced by selecting a single subsurface structure in advance was improved by considering subsurface structural uncertainties. Additionally, time sections obtained from separate a seismic reflection survey conducted in the same area revealed two potential structural boundaries that could represent the plate boundary. By converting these time sections into depth sections using the velocity structure ensemble, it became possible to evaluate the uncertainty of the depth of these structural boundaries. The hypocenter determination results were found to be consistent with these structural boundaries under uncertainty.

Because the publication of the paper on this research [Agata et al. (2025)] came after the acceptance of this manuscript, the readers are referred to that paper for detailed information. However, this study serves as an example demonstrating the importance of considering how uncertainties in subsurface structures propagate into hypocenter determination and other source inversion processes, as discussed in Sections 2 and 3.

## 5. Summary and Future Prospects

In earthquake source inversion, which aims to understand diverse fault activities on earthquake faults using seismic observation data, uncertainties in velocity structure models are typically not considered. In this article, I outlined the author's previous efforts to quantitatively evaluate the uncertainty in velocity structure models and appropriately incorporate its propagation into earthquake source inversion. Firstly, in Section 2, I explained a Bayesian multi-model source inversion method that can incorporate velocity structure model uncertainties when these uncertainties are known as probability distributions, represented as ensembles. Secondly, in Section 3, I discussed a Bayesian traveltime tomography method utilizing PINN to quantify the uncertainty in velocity structure models. Finally, in Section 4, I introduced recent efforts that integrate both approaches and apply them to hypocenter determination in the Nankai Trough region.

The example discussed in Section 4 is a very simple one, focusing on a two-dimensional P-wave velocity structure model and hypocenter determination. In the future, I aim to extend this approach to three-dimensional velocity structure models by performing similar ensemble estimations and integrating them with Bayesian multi-model source inversion using crustal deformation data and seismic waveform data. In the current example, the authors conducted our own velocity structure modeling for the region of interest for hypocenter determination. However, in the future, it would be more efficient to develop ensemble velocity structure models for each seismogenic zone, such as subduction zones, in advance, and use them when performing source inversion. Additionally, by pre-training PINN models to learn the responses at observation points for various source parameters under each velocity structure sample, as described in Section, we can perform Bayesian multi-model source inversion more rapidly. Aiming at such applications in the future, the authors have been also working on developing a rapid hypocenter determination method using PINN models that have learned travel times for various sources based on given velocity structure models of the Nankai Trough region [Agata et al. (under review)]. Such an approach is also applicable to seismic waveform modeling [e.g., Ding et al. (2023); Ren et al. (2024)] and crustal deformation modeling [Okazaki et al. (2022, 2024)] using PINNs. The field of scientific machine learning (SciML) [Baker et al. (2019)], which is represented by PINN and aims to solve scientific problems through the synergistic effects of machine learning and scientific computing (physical laws), has been gaining attention in recent years. By utilizing other SciML techniques than PINN, such as operator learning [Lu et al. (2021); Li et al. (2021)], which can more flexibly learn equation solutions under various analytical conditions, we can take even better approaches for accelerating the Bayesian multi model estimations using pre-trained deep learning models. Thus, I expect that source inversion, which takes into account the propagation of velocity structure model uncertainties quantified by Bayesian estimation, will become much more feasible compared to traditional methods with the application of new machine learning technologies like SciML.

## Acknowledgements

This article was written at the recommendation of Dr. Yukitoshi Fukahata, the Editor-in-Chief of *Zisin* (*Journal of the Seismological Society of Japan. 2nd ser.*) to commemorate receiving the 2024 Young Scientist Award from the Seismological Society of Japan. I would like to express my gratitude to Dr. Yuji Yagi, Dr. Kazuro Hirahara, Dr. Gou Fujie, and Dr. Takane Hori for recommending me for the award. The comments from Dr. Kazuhiro Somei, the editorial committee member, and Dr. Tomohisa Okazaki, the reviewer, were very helpful in improving the manuscript.

## References

Agata, R., 2020, Introduction of covariance components in slip inversion of geodetic data following a non-uniform spatial distribution and application to slip deficit rate estimation in the Nankai Trough subduction zone, Geophys. J. Int., 221, 1832–1844.

Agata, R., A. Kasahara, and Y. Yagi, 2021, A Bayesian inference framework for fault slip distributions based on ensemble modelling of the uncertainty of underground structure: with a focus on uncertain fault dip, Geophys. J. Int., 225, 1392–1411.

Agata, R., K. Shiraishi, and G. Fujie, 2023, Bayesian Seismic Tomography Based on Velocity-Space Stein Variational Gradient Descent for Physics-Informed Neural Network, IEEE Trans. Geosci. Remote Sens., 61, 1-17, doi:10.1109/TGRS.2023.3295414.

Agata, R., K. Shiraishi, and G. Fujie, 2025, Physics-informed deep learning quantifies propagated uncertainty in seismic structure and hypocenter determination, *Sci. Rep.*, 15(1), 1846, https://doi.org/10.1038/s41598-024-84995-9.

Agata, R., R. Nakata, A. Kasahara, Y. Yagi, Y. Seshimo, S. Yoshioka, and T. Iinuma, 2022, Bayesian multi-model estimation of fault slip distribution for slow slip events in southwest Japan: Effects of prior constraints and uncertain underground structure, J. Geophys. Res.: Solid Earth, 127, 8, e2021JB023712.

Agata, R., S. Baba, A. Nakanishi, Y. Nakamura, under review, HypoNet Nankai: Rapid hypocenter determination tool for the Nankai Trough subduction zone using physics-informed neural networks. arXiv preprint arXiv:2411.04667.

Agata, R., T. Ichimura, T. Hori, K. Hirahara, C. Hashimoto, and M. Hori, 2018, An adjoint-based simultaneous estimation method of the asthenosphere's viscosity and afterslip using a fast and scalable finite-element adjoint solver, Geophys. J. Int., 213, 461–474.

Akaike, T., 1980, Likelihood and Bayes procedure, Bayesian Statistics, 143–166.

Aoi, S., Y. Asano, T. Kunugi, T. Kimura, K. Uehira, N. Takahashi, H. Ueda, K. Shiomi, T. Matsumoto, and H. Fujiwara, 2020, MOWLAS: NIED observation network for earthquake, tsunami, and volcano, Earth, Planets and Space, 72(1), 1–31, Springer.

Baker, N., F. Alexander, T. Bremer, A. Hagberg, Y. Kevrekidis, H. Najm, M. Parashar, A. Patra, J. Sethian, S. Wild, K. Willcox, and S. Lee, 2019, Workshop Report on Basic Research Needs for Scientific Machine Learning: Core Technologies for Artificial Intelligence.

Bodin, T. and M. Sambridge, 2009, Seismic tomography with the reversible jump algorithm, Geophys. J. Int., 178, 3, 1411–1436.

Chen, Y., S. A. L. de Ridder, S. Rost, Z. Guo, X. Wu, and Y. Chen, 2022, Eikonal Tomography With Physics-Informed Neural Networks: Rayleigh Wave Phase Velocity in the Northeastern Margin of the Tibetan Plateau, Geophys. Res. Lett., 49(21), e2022GL099053.

Ding, Y., S. Chen, X. Li, S. Wang, S. Luan, and H. Sun, 2023, Self-adaptive physics-driven deep learning for seismic wave modeling in complex topography, Eng. Appl. Artif. Intell., 123, 106425.

Duane, S., A. D. Kennedy, B. J. Pendleton, and D. Roweth, 1987, Hybrid monte carlo, Phys. Lett. B, 195(2), 216–222.

Duputel, Z., P. S. Agram, M. Simons, S. E. Minson, and J. L. Beck, 2014, Accounting for prediction uncertainty when inferring subsurface fault slip, Geophys. J. Int., 197, 464-482.

Dutta, R., S. Jónsson, and H. Vasyura-Bathke, 2021, Simultaneous Bayesian estimation of non-planar fault geometry and spatially-variable slip, J. Geophys. Res. Solid Earth, 126, 7, e2020JB020441.

Fukahata, Y., and T. J. Wright, 2008, A non-linear geodetic data inversion using ABIC for slip distribution on a fault with an unknown dip angle, Geophys. J. Int., 173, 353–364.

Fukuda, J., and K. M. Johnson, 2008, A fully Bayesian inversion for spatial distribution of fault slip with objective smoothing, Bull. Seismol. Soc. Am., 98, 1128–1146.

Fukuda, J., and K. M. Johnson, 2010, Mixed linear–non-linear inversion of crustal deformation data: Bayesian inference of model, weighting and regularization parameters, Geophys. J. Int., 181, 1441–1458.

Gesret, A., Desassis, N., Noble, M., Romary, T., and Maisons, C., 2015, Propagation of the velocity model uncertainties to the seismic event location, Geophys. J. Int., 200(1), 52–66.

Green, P. J., 1995, Reversible jump Markov chain Monte Carlo computation and Bayesian model determination, Biometrika, 82(4), 711–732.

Gombert, B., Z. Duputel, R. Jolivet, M. Simons, J. Jiang, C. Liang, E. J. Fielding, and L. Rivera, 2018, Strain budget of the Ecuador–Colombia subduction zone: A stochastic view, Earth Planet. Sci. Lett., 498, 288–299.

Gou, R., Y. Zhang, X. Zhu, and J. Gao, 2023, Bayesian physics-informed neural networks for the subsurface tomography based on the eikonal equation, IEEE Trans. Geosci. Remote Sens., 61, 1–12.Hallo, M., and F. Gallovič, 2016, Fast and cheap approximation of Green function uncertainty for waveform-based earthquake source inversions, Geophys. J. Int., 207, 1012–1029.

Hayes, G. P., G. L. Moore, D. E. Portner, M. Hearne, H. Flamme, M. Furtney, and G. M. Smoczyk, 2018, Slab2, a comprehensive subduction zone geometry model, Science, 362, 6410, 58–61.

Hirata, N., and M. Matsu'ura, 1987, Maximum-likelihood estimation of hypocenter with origin time eliminated using nonlinear inversion technique, Phys. Earth Planet. Inter., 47, 50–61.

Hirose, H., Y. Asano, K. Obara, T. Kimura, T. Matsuzawa, S. Tanaka, and T. Maeda, 2010, Slow earthquakes linked along dip in the Nankai subduction zone, Science, 330, 6010, 1502.

Iwasaki, T., H. Sato, T. Ishiyama, M. Shinohara, and A. Hashima, 2015, Fundamental structure model of island arcs and subducted plates in and around Japan, in AGU Fall Meeting Abstracts, 2015, T31B–2878.

Kaneda, Y., K. Kawaguchi, E. Araki, H. Matsumoto, T. Nakamura, S. Kamiya, K. Ariyoshi, T. Hori, T. Baba, and N. Takahashi, 2015, Development and application of an advanced ocean floor network system for megathrust earthquakes and tsunamis, in Seafloor observatories, 643–662, Springer.

Kano, M., Aso, N., Matsuzawa, T., Ide, S., Annoura, S., Arai, R., Baba, S., Bostock, M., Chao, K., Heki, K., Itaba, S., Ito, Y., Kamaya, N., Maeda, T., Maury, J., Nakamura, M., Nishimura, T., Obana, K., Ohta, K., Poiata, N., Rousset, B., Sugioka, H., Takagi, R., Takahashi, T., Takeo, A., Tu, Y., Uchida, N., Yamashita, Y., and Obara, K., 2018, Development of a Slow Earthquake Database, Seismol. Res. Lett., 89(4), 1566–1575.Koketsu, K., H. Miyake, Y. Tanaka, et al., 2009, A proposal for a standard procedure of modeling 3-D velocity structures and its application to the Tokyo metropolitan area, Japan, Tectonophysics, 472, 1–4, 290–300.

Koketsu, K., H. Miyake, and H. Suzuki, 2012, Japan integrated velocity structure model version 1, in Proceedings of the 15th World Conference on Earthquake Engineering, 1773, Lisbon.

Korenaga, J., W. S. Holbrook, G. M. Kent, P. B. Kelemen, R. S. Detrick, H.-C. Larsen, J. R. Hopper, and T. Dahl-Jensen, 2000, Crustal structure of the southeast Greenland margin from joint refraction and reflection seismic tomography, J. Geophys. Res., 105, B9, 21591-21614.

Li, Z., N. B. Kovachki, K. Azizzadenesheli, K. Bhattacharya, A. Stuart, A. Anandkumar, et al., 2020, Fourier Neural Operator for Parametric Partial Differential Equations, International Conference on Learning Representations.

Linka, K., A. Schäfer, X. Meng, Z. Zou, G.E. Karniadakis, and E. Kuhl, 2022, Bayesian physics-informed neural networks for real-world nonlinear dynamical systems, Comput. Methods Appl. Mech. Eng., 402, 115346.

Lindsey, E. O. and Y. Fialko, 2013, Geodetic slip rates in the southern San Andreas Fault system: Effects of elastic heterogeneity and fault geometry, J. Geophys. Res. Solid Earth, 118(2), 689–697.

Liu, Q., and D. Wang, 2016, Stein Variational Gradient Descent: A General Purpose Bayesian Inference Algorithm, Adv. Neural Inf. Process. Syst., 29.

Lomax, A., J. Virieux, P. Volant, and C. Berge-Thierry, 2000, Probabilistic earthquake location in 3D and layered models: Introduction of a Metropolis-Gibbs method and comparison with linear locations, in Advances in Seismic Event Location, 101–134, Springer.

Lu, L., P. Jin, G. Pang, Z. Zhang, and G. E. Karniadakis, 2021, Learning nonlinear operators via DeepONet based on the universal approximation theorem of operators, Nature Machine Intelligence, 3(3), 218–229, Nature Publishing Group UK London.

Maeda, T. and K. Obara, 2009, Spatiotemporal distribution of seismic energy radiation from low-frequency tremor in western Shikoku, Japan, J. Geophys. Res., 114, B10.

Minson, S. E., M. Simons, J. L. Beck, F. Ortega, J. Jiang, S. E. Owen, A. W. Moore, A. Inbal, and A. Sladen, 2014, Bayesian inversion for finite fault earthquake source models—II: the 2011 great Tohoku-oki, Japan earthquake, Geophys. J. Int., 198, 922–940.

Miyazaki, S. and Hatanaka, Y., 1998, The outlines of the GEONET (in Japanese), Meteorol. Res. Note, 192, 105–131.

Nakanishi, A., N. Takahashi, Y. Yamamoto, T. Takahashi, S. O. Citak, T. Nakamura, K. Obana, S. Kodaira, and Y. Kaneda, 2018, Three-dimensional plate geometry and P-wave velocity models of the subduction zone in SW Japan: Implications for seismogenesis, in Geology and Tectonics of Subduction Zones: A Tribute to Gaku Kimura, 534, 69.

Nakano, M., M. Hyodo, A. Nakanishi, M. Yamashita, T. Hori, S. Kamiya, K. Suzuki, T. Tonegawa, S. Kodaira, N. Takahashi, and Y. Kaneda, 2018, The 2016 Mw 5.9 earthquake off the southeastern coast of Mie Prefecture as an indicator of preparatory processes of the next Nankai Trough megathrust earthquake, Prog. Earth Planet. Sci., 5(1), 1–17.

Nakao, K., T. Ichimura, K. Fujita, T. Hori, T. Kobayashi, and H. Munekane, 2024, Massively parallel Bayesian estimation with Sequential Monte Carlo sampling for simultaneous estimation of earthquake fault geometry and slip distribution, J. Comput. Sci., 81, 102372.

Obara, K., S. Tanaka, T. Maeda, and T. Matsuzawa, 2010, Depth-dependent activity of non-volcanic tremor in southwest Japan, Geophys. Res. Lett., 37, L13306.

Okazaki, T., T. Ito, K. Hirahara, and N. Ueda, 2022, Physics-informed deep learning approach for modeling crustal deformation, Nat. Commun., 13(1), 7092.

Okazaki, T., K. Hirahara, and N. Ueda, 2024, Fault geometry invariance and dislocation potential in antiplane crustal deformation: physics-informed simultaneous solutions, Prog. Earth Planet. Sci., 11(1), 52.

Piana Agostinetti, N., G. Giacomuzzi, and A. Malinverno, 2015, Local three-dimensional earthquake tomography by trans-dimensional Monte Carlo sampling, Geophys. J. Int., 201, 3, 1598–1617.

Poliannikov, O. V., M. Prange, A. E. Malcolm, and H. Djikpesse, 2014, Joint location of microseismic events in the presence of velocity uncertainty, Geophysics, 79, 6, KS51–KS60.

Puel, S., T. W. Becker, U. Villa, O. Ghattas, and D. Liu, 2024, Volcanic arc rigidity variations illuminated by coseismic deformation of the 2011 Tohoku-oki M9, Science Advances, 10, 23, eadl4264.

Raftery, A. E., D. Madigan, and J. A. Hoeting, 1997, Bayesian model averaging for linear regression models, J. Am. Stat. Assoc., 92, 437, 179–191.

Ragon, T., A. Sladen, and M. Simons, 2019, Accounting for uncertain fault geometry in earthquake source inversions—II: Application to the M_w 6.2 Amatrice earthquake, central Italy, Geophys. J. Int., 218, 1, 689–707.

Raissi, M., P. Perdikaris, and G. E. Karniadakis, 2019, Physics-informed neural networks: A deep learning framework for solving forward and inverse problems involving nonlinear partial differential equations, J. Comput. Phys., 378, 686–707.

Ren, P., C. Rao, S. Chen, J.-X. Wang, H. Sun, and Y. Liu, 2024, SeismicNet: Physics-informed neural networks for seismic wave modeling in semi-infinite domain, Comput. Phys. Commun., 295, 109010.

Ryberg, T., C. Haberland, B. Wawerzinek, M. Stiller, K. Bauer, A. Zanetti, L. Ziberna, G. Hetényi, O. Münstener, M. M. Weber, and others, 2023, 3-D imaging of the Balmuccia peridotite body (Ivrea–Verbano zone, NW-Italy) using controlled source seismic data, Geophys. J. Int., 234, 3, 1985–1998.

Ryberg, T., and Ch. Haberland, 2018, Bayesian inversion of refraction seismic traveltime data, Geophys. J. Int., 212, 3, 1645–1656.

Seshimo, Y., and S. Yoshioka, 2021, Spatiotemporal slip distributions associated with the 2018-2019 Bungo Channel long-term slow slip event inverted from GNSS data, Scientific Reports.

Shimizu, K., Y. Yagi, R. Okuwaki, and Y. Fukahata, 2021, Construction of fault geometry by finite-fault inversion of teleseismic data, Geophys. J. Int., 224, 2, 1003–1014.

Smith, J. D., K. Azizzadenesheli, and Z. E. Ross, 2021, EikoNet: Solving the Eikonal Equation With Deep Neural Networks, IEEE Trans. Geosci. Remote Sens., 59(12), 10685–10696, doi: 10.1109/TGRS.2020.3039165.

Tarantola, A., and B. Valette, 1982, Inverse problems = quest for information, J. Geophys., 50, 159–170.

Tarantola, A., 2005, Inverse problem theory and methods for model parameter estimation, SIAM, Philadelphia, PA.

Tebaldi, C., and R. Knutti, 2007, The use of the multi-model ensemble in probabilistic climate projections, Philos. Trans. R. Soc. A Math. Phys. Eng. Sci., 365, 1857, 2053–2075.

Waheed, U. B., E. Haghighat, T. Alkhalifah, C. Song, and Q. Hao, 2021a, PINNeik: Eikonal solution using physics-informed neural networks, Comput. Geosci., 155, 104833.

Waheed, U. B., T. Alkhalifah, E. Haghighat, C. Song, and J. Virieux, 2021b, PINNtomo: Seismic tomography using physics-informed neural networks, arXiv preprint arXiv:2104.01588.

Wallace, L. M., E. Araki, D. Saffer, X. Wang, A. Roesner, A. Kopf, A. Nakanishi, W. Power, R. Kobayashi, C. Kinoshita, et al., 2016, Near-field observations of an offshore Mw 6.0 earthquake from an integrated seafloor and subseafloor monitoring network at the Nankai Trough, southwest Japan, J. Geophys. Res. Solid Earth, 121(11), 8338–8351.

Wang, R., F. L. Martín, and F. Roth, 2003, Computation of deformation induced by earthquakes in a multi-layered elastic crust—FORTRAN programs EDGRN/EDCMP, Comput. Geosci., 29, 2, 195–207.

Wang, Z., T. Ren, J. Zhu, and B. Zhang, 2019, Function Space Particle Optimization for Bayesian Neural Networks, in Proc. Int. Conf. Learn. Represent.

Yagi, Y. and Y. Fukahata, 2011, Introduction of uncertainty of Green's function into waveform inversion for seismic source processes, Geophys. J. Int., 186, 711-720.

Yagi, Y., and Y. Fukahata, 2008, Importance of covariance components in inversion analyses of densely sampled observed data: an application to waveform data inversion for seismic source processes, Geophys. J. Int., 175, 215–221, doi:10.1111/j.1365-246X.2008.03929.x.

Yang, L., X. Meng, and G. E. Karniadakis, 2021, B-PINNs: Bayesian physics-informed neural networks for forward and inverse PDE problems with noisy data, J. Comput. Phys., 425, 109913.

Yoshioka, S., Y. Matsuoka, and S. Ide, 2015, Spatiotemporal slip distributions of three long-term slow slip events beneath the Bungo Channel, southwest Japan, inferred from inversion analyses of GPS data, Geophys. J. Int., 201, 3, 1437–1455.

Zhang, J. and M. N. Toksöz, 1998, Nonlinear refraction traveltime tomography, Geophysics, 63, 5, 1726–1737.

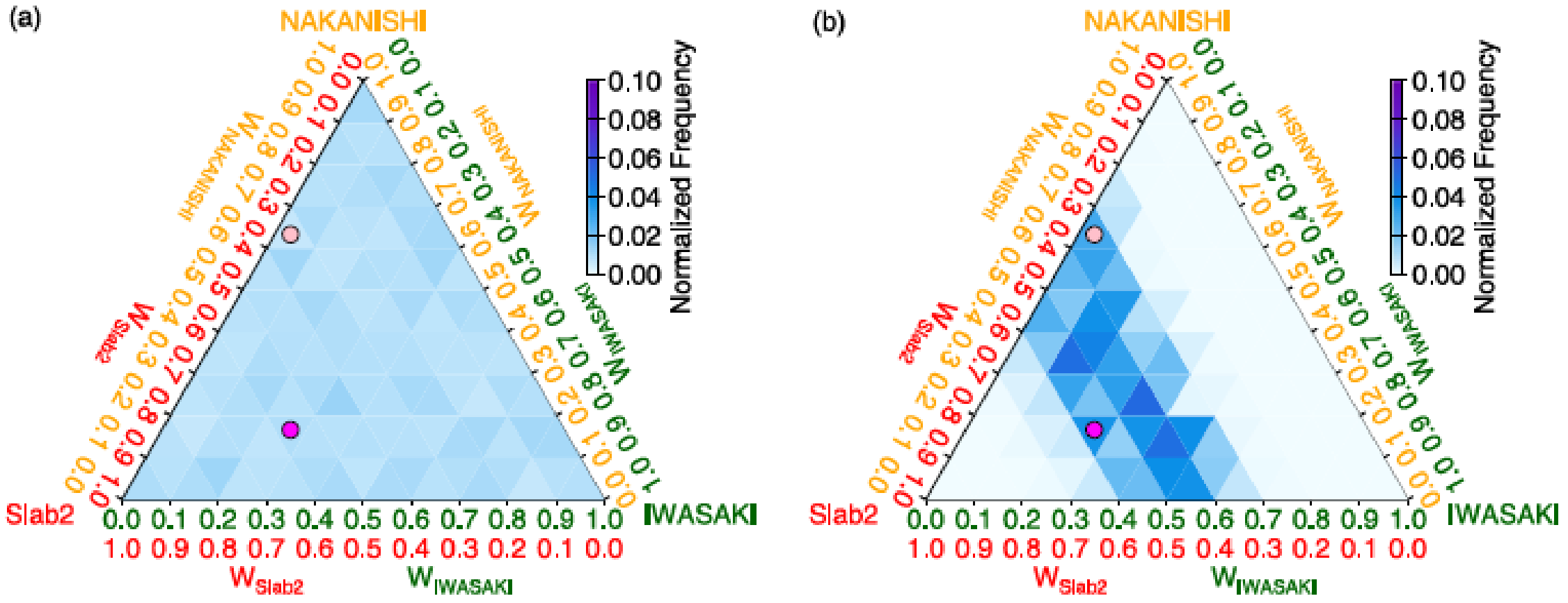

Fig. 1. Comparison of ternary plots showing samples from (a) the prior and (b) the posterior probability density functions (PDFs) for the 2010 estimation of the plate boundary geometry model for the L-SSE occurred in 2010, described in Agata et al. (2022). Pink and magenta circles mark locations designated for further visualizations, which are not included in this article. Reproduced with permission from Agata et al., Journal of Geophysical Research: Solid Earth, 2022. Copyright © 2022, John Wiley and Sons.

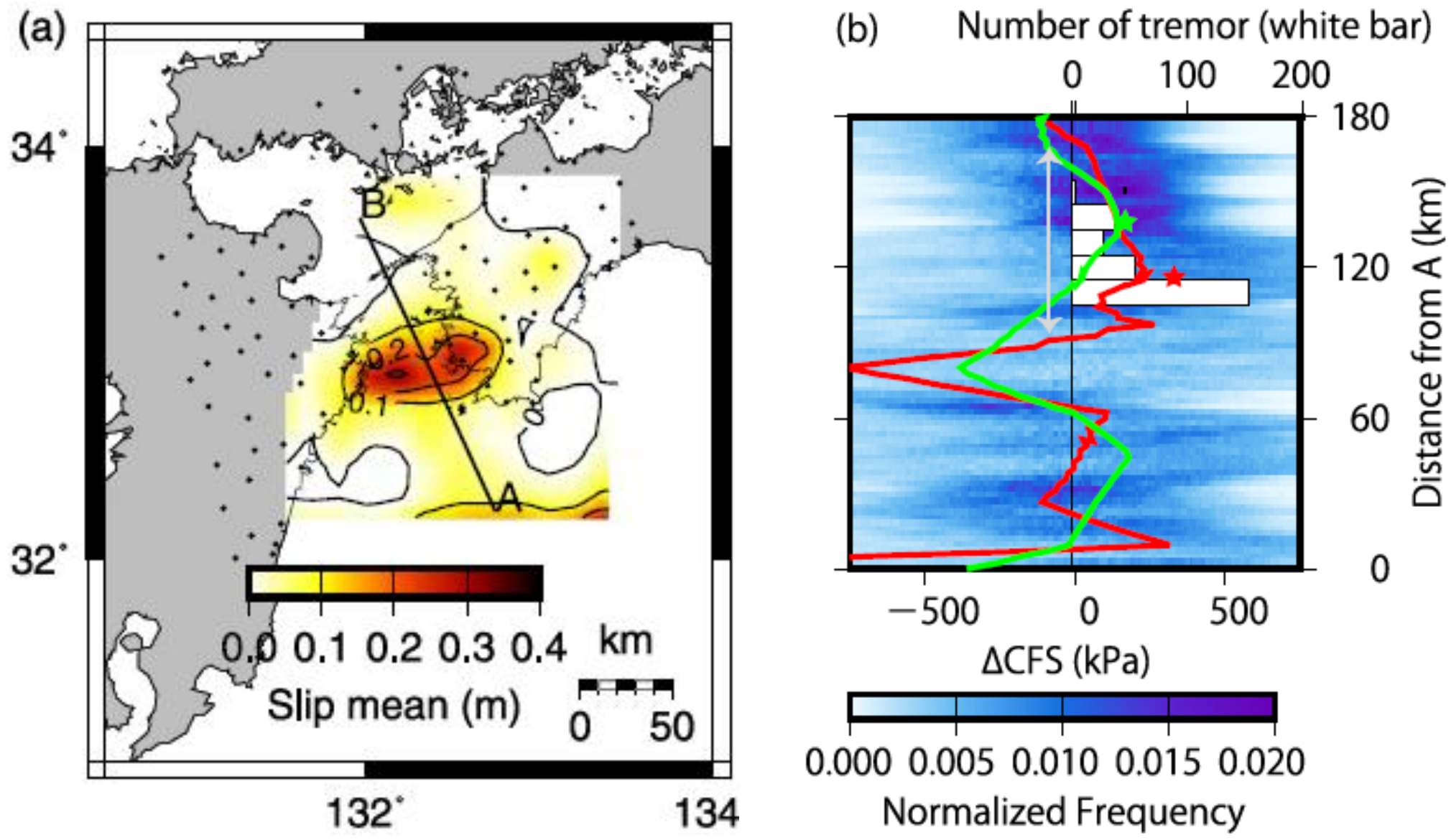

Fig. 2. (a) The estimated mean model of posterior probability density function (PDF) for slip distribution for the L-SSE occurred in 2010 obtained by Agata et al. (2022). The dots indicate GEONET [Miyazaki and Hatanaka (1998)] observation points for crustal deformation used in the study. (b) The correspondence between ΔCFS calculated using the estimated mean model and the tremor distribution on the A-B line profile. The blue color map denotes the frequencies of the ΔCFS values for the posterior PDF. The red and green lines denote the distribution of the mean of ΔCFS calculated based on the posterior PDF of the slip distribution estimated by the proposed method and that calculated based on the slip distribution of the smoothing model, respectively. The location of the peak of the positive value of the mean ΔCFS in the down-dip side of the channel for BMMFSE and the smoothing model are denoted by red and green stars, respectively. The white bars denote the number of tremors during the L-SSE period in the area within 5 km from the line in the direction perpendicular to it. Reproduced with permission from Agata et al., Journal of Geophysical Research:

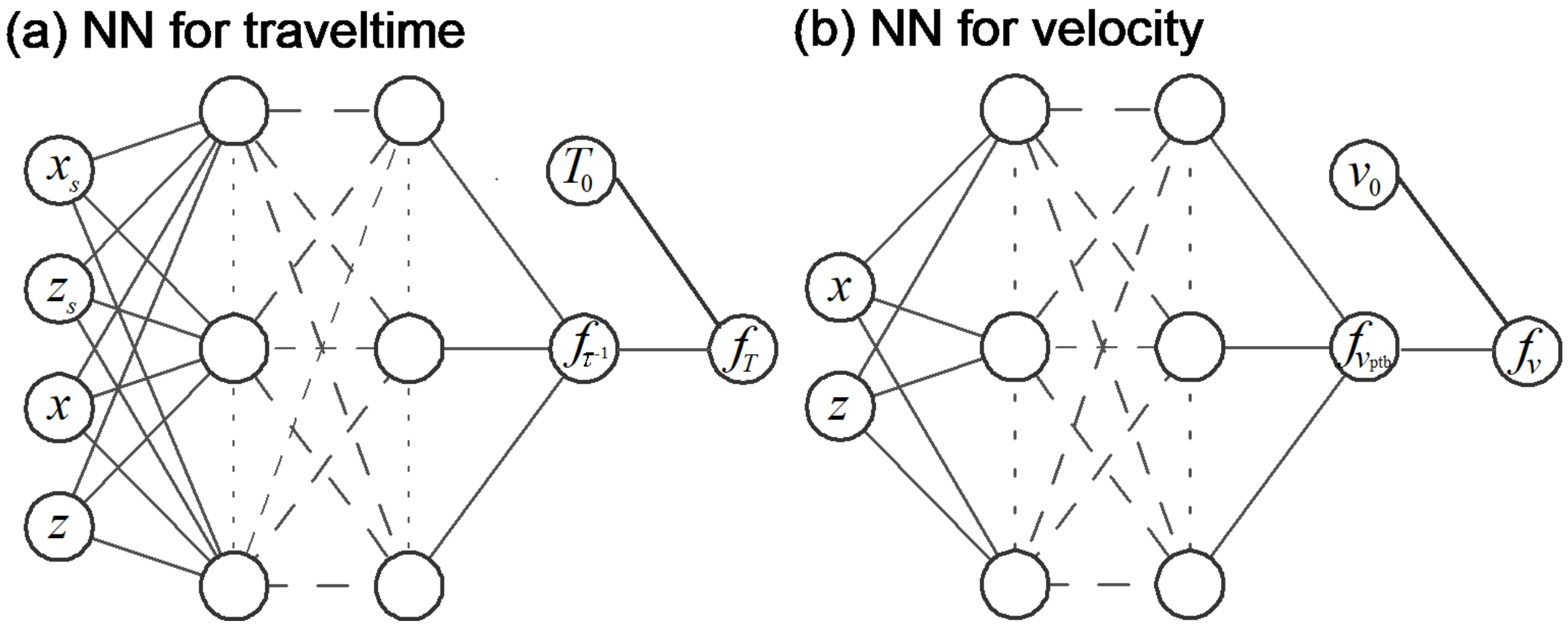

Fig. 3. (a) A schematic of the neural network (NN) employed for function approximation of travel time. (b) That for velocity structure. $f_T$ and $f_v$ are not a direct output of NNs, but it undergoes additional manipulations (see the main text and Agata et al. (2023))

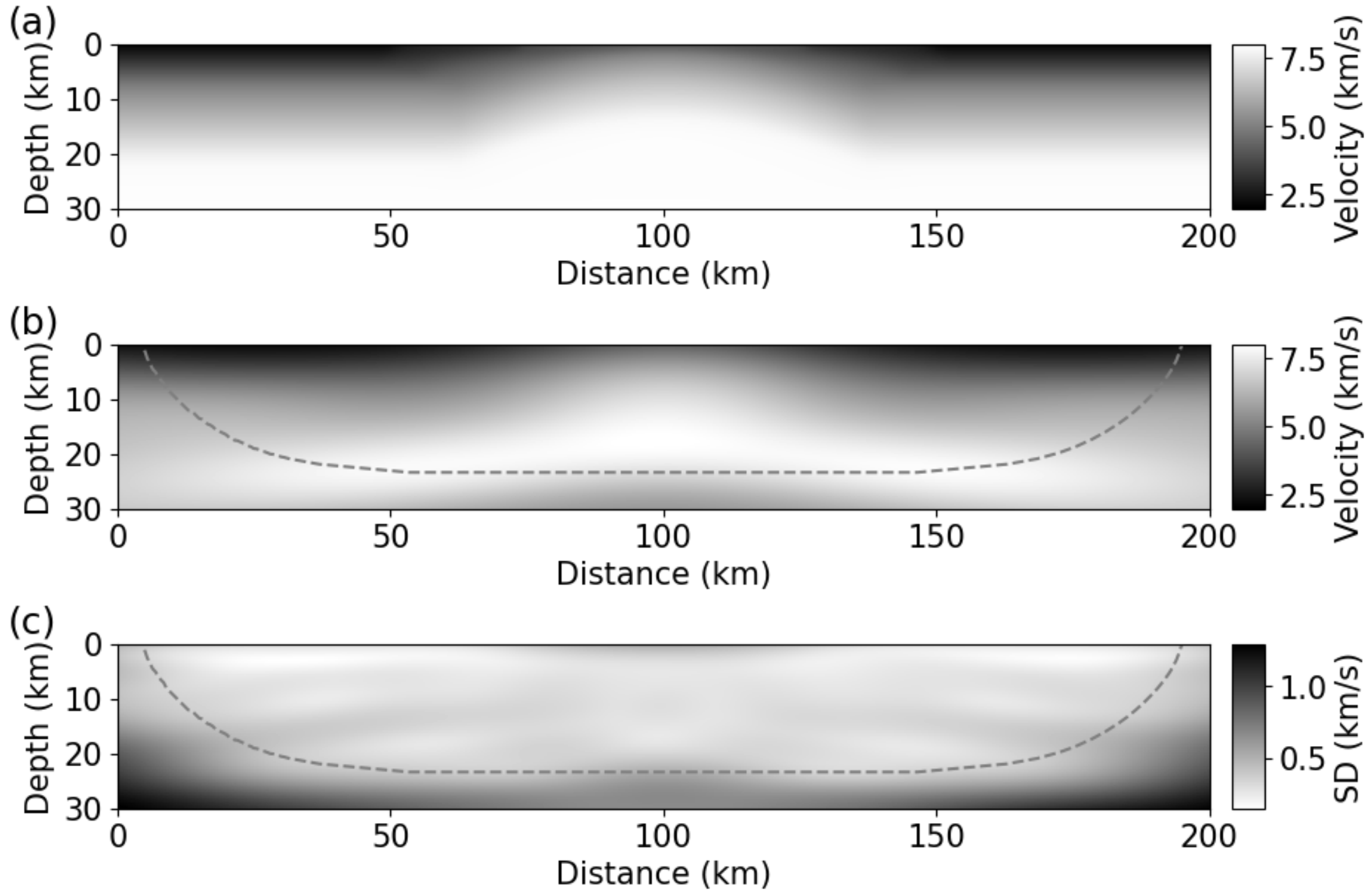

Fig. 4. (a) The true model, (b) estimated mean model and (c) standard deviation for the ensemble velocity structure model that are used and obtained through 2D PINN-based Bayesian traveltime tomography of Agata et al. (2023).

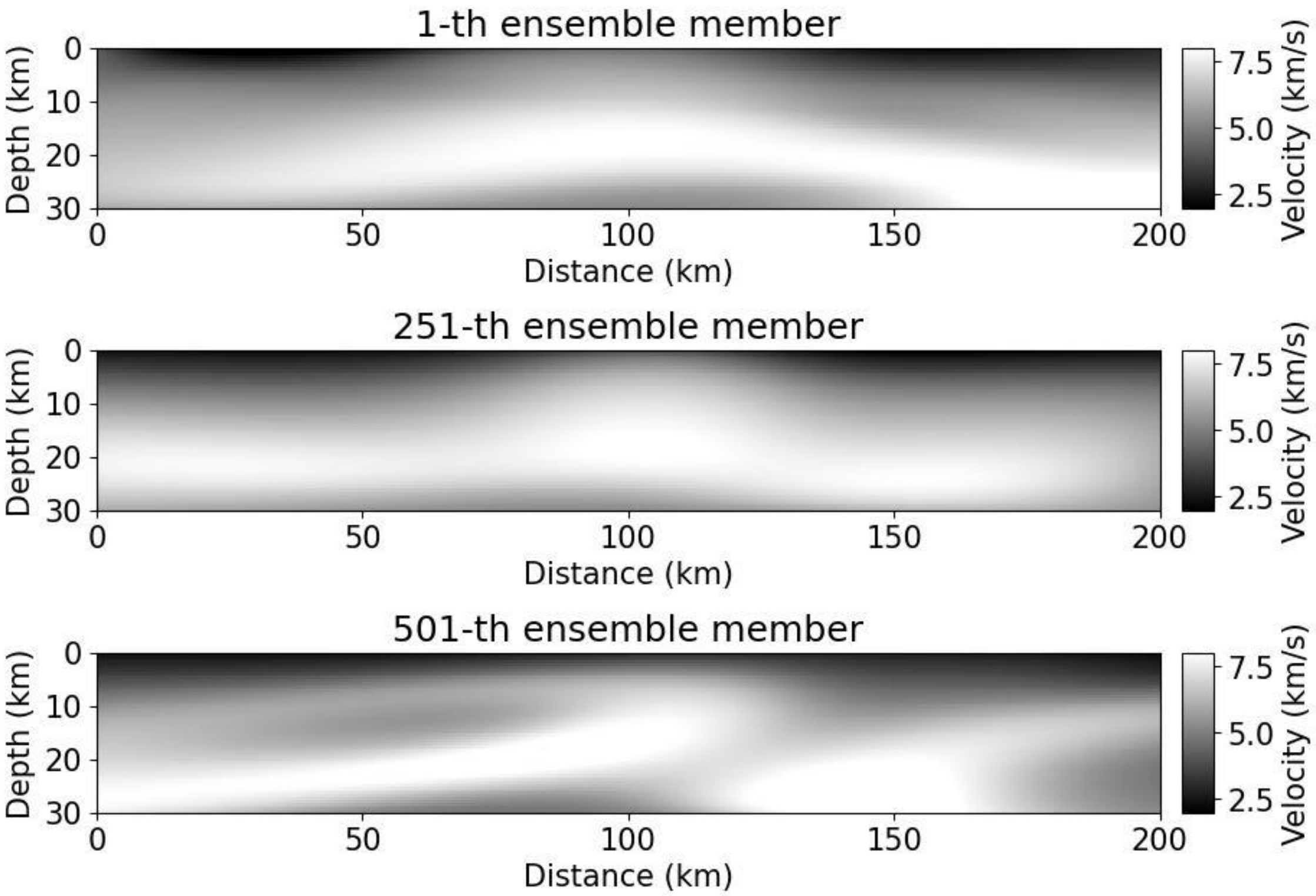

Fig. 5. Examples of velocity structure samples obtained through 2D PINN-based Bayesian traveltime tomography of Agata et al. (2023).

Table 1. Summary of practical issues in B-PINN based on Hamiltonian Monte Carlo (HMC) and solutions provided by the proposed method based on particle-based variational inference (ParVI) in function space.

| | **B-PINN based on HMC** | **B-PINN based on function-space ParVI (proposed)** | **Related technique in the proposed method** |
|---|---|---|---|
| **Multimodality of posterior PDF** | Large due to targeting posterior PDF in the weight parameter space | Comparatively small due to targeting posterior PDF in the function (physical) space | Estimation in the function space |
| **Parallelism** | Low due to the need for numerous sequential trials | High owing to high independence of operations for each particle | ParVI |
| **Mini-batch training** | Not applicable | Applicable | ParVI |
| **Prior PDF** | Tend to lack physical meaning because it is defined in the weight parameter space | Possess physical meaning because it is defined in the function (physical) space | Estimation in the function space |